\documentstyle[aps,prl,twocolumn]{revtex} 

\def\be{\begin{equation}}
\def\ee{\end{equation}}
\def\bea{\begin{eqnarray}}
\def\eea{\end{eqnarray}}
\def\bma{\begin{mathletters}}
\def\ema{\end{mathletters}}

\begin{document}


\title{Quantum gates with "hot" trapped ions}

\author{J. F. Poyatos$^{(1,2)}$, J.I. Cirac$^{(1,2)}$, and P. Zoller$^{(2)}$}

\address{
(1) Departamento de F\'{\i}sica Aplicada, Universidad de Castilla--La Mancha,
13071 Ciudad Real, Spain\\
(2) Institut f\"ur Theoretische Physik, Universit\"at Innsbruck,
Technikerstrasse 25, A--6020 Innsbruck, Austria.
}

\date{\today}

\maketitle

\begin{abstract}
We propose a scheme to perform a fundamental two--qubit gate between
two trapped ions using ideas from atom interferometry. 
As opposed to the scheme considered by J. I. Cirac and P. Zoller, 
Phys. Rev. Lett. 74, 4091 (1995), it does
not require laser cooling to the motional ground state.
\end{abstract}

\pacs{PACS: 03.65.Bz, 32.80.Pj, 89.80+h}



Quantum computation allows the development of polynomial-time algorithms
for computational problems such a prime factoring, which have previously
been viewed as intractable~\cite{St97}. This has motivated studies into
the feasibility of actual implementation of quantum computers in
physical systems~\cite{Qc}. The task of designing a QC is equivalent to
finding a physical realization of quantum gates between a set of qubits,
where a qubit refers to a two--level system $\{ |0\rangle, |1\rangle
\}$. Any operation can be decomposed into rotations on a single qubit
and a universal two-bit gate, e.g. a gate defined by 
\begin{equation}
\label{gate}
\hat{C}_{12}: |\epsilon_1 \rangle |\epsilon_2\rangle \rightarrow
|\epsilon_1\rangle |(1+\epsilon_1) \oplus \epsilon_2\rangle
\quad (\epsilon_{1,2}=0,1). 
\end{equation}
Implementation of a quantum computer requires precise control of the
Hamiltonian operations and a high degree of coherence. Achieving the
conditions for quantum computation is extremely demanding, and only a
few systems have been identified as possible candidates to build small
scale models in the lab~\cite{St97,Qc,Ci95,NMR}. One of the most promising
examples is a string of cold ions stored in a linear trap~\cite{Ci95}. In this ion
trap quantum computer qubits are stored in long--lived internal atomic
ground states. Single bit operations in this system are accomplished by
directing different laser beams to each of the ions, and a fundamental
gate operation is implemented by exciting the collective quantized
motion of the ions with lasers, where the exchange of phonons serves as
a data bus to transfer quantum information between the
qubits. Prospects of building small ion trap quantum
computers in the near future are good, but the question remains whether
stringent requirements in implementing two bit quantum gates can be
relaxed: while decoherence of the qubit stored in the atomic ground
state is not an issue, cooling of ions to the vibrational ground state
to prepare a pure initial state for the collective phonon mode remains a
challenge~\cite{Di89}. The question arises whether quantum gates can be
performed starting from thermal or mixed states of phonon modes. The
task of implementing quantum computing in "hot" systems seems
particularly timely in view of the recent interest in NMR quantum
computing \cite{NMR}, where quantum states are stored as pseudo-pure
states in an ensemble of "hot" spins.

In this Letter we will discuss the implementation of a universal
two--bit gate between two ions in a linear trap with the phonon modes
initially in a thermal (mixed) state. The novel concept behind the gate
operation is to implement conditional dynamics based on {\em atom
interferometry of two entangled atoms}~\cite{Po96,Ao}. The two-bit gate
operation proceeds as follows: ion 2 is kicked left or right depending
on its internal state with laser light (see Fig.~1(a) at $t=0$). Thus the
ion 1 will experience a kick via the Coulomb repulsion conditional to
the internal state of the first ion. The corresponding wave packet will
evolve into a superposition of two spatial wavepackets which are
entangled to the internal state of the control ion. Provided the spatial
splitting of the wave packet of the second ion is sufficiently large (at
time $t_0$ in Fig.~1(a)), we can manipulate the internal state of the
target atom depending on its spatial position, i.e depending on the
state of the first ion, and thus implement a gate operation on the
qubits. With time these atomic wave packets will oscillate in the trap,
and with a proper sequence of laser pulses this momentum transfered to
the two atoms can be undone to restore the original motional state of
the ion (Fig.~1(a) at time $t_g$). The motional state of the atom will thus
factorize from the internal atomic state before and after the gate,
independent of being a mixed or a pure state. 

In order to analyze the scheme, we consider two ions confined in a
linear trap~\cite{Ion}. We will assume that the motion of the ions is
frozen with respect to the $y$ and $z$ axis. The Hamiltonian describing
this situation is
\be
H = \frac{p_1^2}{2m} +  \frac{p_2^2}{2m} + V(x_1) + V(x_2) + 
\frac{e^2}{4\pi\epsilon_0|x_1-x_2|} ,
\ee
where $V$ denotes the external confining potential along the $x$ axis,
which we assume to be symmetric $V(x)=V(-x)$. We denote by $x_e$ the
separation of the ions at the equilibrium position. Using the
center--off--mass and the relative coordinates, 
\bma
\bea
x_{c} &=& (x_1 + x_2)/2, \quad\quad x_r = x_1 - x_2 - x_e,\\
p_{c} &=&  p_1 + p_2,    \quad\quad p_r =(p_1 - p_2)/2,
\eea
\ema
and expanding around the equilibrium position $x_c=x_r=0$ up to second
order, we obtain 
\be
\label{Htot}
H = H_{\rm ho} + V_{\rm cor}(x_c,x_r),
\ee
where
\be
\label{Hharm}
H_{\rm ho} = \frac{p_c^2}{2m_c} +  \frac{p_r^2}{2m_r} + 
\frac{1}{2} m_c \nu_c^2 x_c^2 + \frac{1}{2} m_r \nu_r^2 x_r^2,
\ee
and $V_{\rm cor}$ denotes the third and higher--order corrections. 
Here $m_c=2m$ and $m_r=m/2$ are the total mass and the reduced mass of
the particles, and $\nu_{c,r}$ the corresponding frequencies in the
harmonic approximation
\be
\nu_c^2= \frac{1}{m_c}\frac{\partial ^2V}{\partial x^2_c}\Big|_{eq} , 
\quad \nu_r^2=\nu_c^2 + \frac{e^2}{\pi\epsilon_0 m x_e^3}.
\ee

In the following we will assume that the trapping potential has been
tuned so that $\nu_r=2 \nu_c$. On the other hand, in order to use the
harmonic approximation to the original Hamiltonian we will neglect the
corrections $V_{\rm cor}$ in (\ref{Hharm}). This requires
$\overline{V_{\rm cor}} t_g \ll 1$, where $\overline{V_{\rm cor}}$
denotes the typical values of these corrections and $t_g=2\pi\nu_c$ the
duration of the gate. A more quantitative analysis of this approximation
will be given below. Under these circumstances, the time evolution of
the ions motion will be strictly periodic, with a period $t_g$; that is,
\be
\label{period}
e^{-i H_{\rm ho} t_g} = 1.
\ee

We will denote by $|0\rangle$ and $|1\rangle$ the two internal states of
atoms which store the quantum information, and by
$|\Psi(0)\rangle=|\Psi_{\rm int}\rangle \otimes |\Psi_{\rm mot}\rangle$
the initial state of the ions. Although we have taken both pure internal
$|\Psi_{\rm int}\rangle$ and motional states $|\Psi_{\rm mot}\rangle$,
our analysis is also valid for mixed initial states. The gate is
performed in three steps:

{\em (i) State dependent kick on atom 2:} A short laser pulse is applied
to atom $2$ with a k--vector pointing along the $x$ direction. The laser
intensity is chosen such that the internal state of the atom undergoes a
flip. The motional state of the ions is also changed by this laser beam.
We write the Hamiltonian for the atom--laser interaction as $H_{\rm
las}=\frac{\Omega}{2} (\sigma^+_2 e^{ikx_2} + \sigma^-_2 e^{-ikx_2})$,
where $\Omega$ is the Rabi frequency, $\sigma^+=|1\rangle\langle
0|=(\sigma^-)^\dagger$ and the subscript denotes from now on which atom
is addressed. The laser is applied for a time $t_{\rm las}=\pi/\Omega
\ll t_g$ (the so--called strong--excitation regime, see~\cite{Po96} for
details), so that the state after this interaction is
\be
|\Psi'(0)\rangle = (\sigma^+_2 e^{ikx_2} + \sigma^-_2 e^{-ikx_2})
|\Psi(0)\rangle.
\ee
According to this expression, if atom $2$ is in the state $|0\rangle$ it
will be transferred to the state $|1\rangle$ and undergo a photon kick
which will push it to the right. If it is initially in $|1\rangle$ it
will be transferred to the state $|0\rangle$ and undergo a photon kick
towards the left [See Fig. 1(a)]. Thus, depending on the internal state of
the atom it will start moving in different directions. Due to the
Coulomb repulsion, atom $1$ will also undergo a different motion
depending on the initial state of atom $2$. That is, the motional state
of the first atom will also split into two wavepackets, $\varphi^R$ and
$\varphi^L$, moving in different directions (right and left). One can
easily calculate the evolution of the distance between the center of these 
wavepackets using the harmonic Hamiltonian (\ref{Hharm})
\be
\label{distance}
d(t) = 2 x_0 \eta \left[ 
\sin(\nu_c t) - \frac{1}{2} \sin(2\nu_c t)
\right] 
\ee
where $x_0=1/(2m\nu_c)^{1/2}$ ($\hbar=1$) is the ground state 
size of a single ion with the  center--of--mass mode frequency  and $\eta=k x_0$ 
the corresponding Lamb--Dicke parameter.
The maximum distance is 
\be
\label{Dx1}
D \equiv d(t_0)=3\sqrt{3}x_0\eta/2,
\ee 
and is produced at $t_0=2\pi/(3\nu_c)$. 

{\em (ii) Conditional flip on atom 1:} After a time $t_0$, a laser beam
is applied to atom $1$, such that it does not suffers a kick
\cite{Note}. The laser is focused on the position $x=(x_e+ D)/2$,
so that it only overlaps with the wavepacket $\varphi^R$, that is, the
one that arises if atom $2$ was in state $|0\rangle$. Adjusting properly
the interaction time $t_1\ll t_g$, it induces a rotation $|0\rangle_1 
\leftrightarrow |1\rangle_1$. The state after this laser interaction will
be
\be
|\Psi(t_0)\rangle = \left[\sigma^x_1 e^{-i H_{\rm ho} t_0}  
\sigma^+_2 e^{ikx_2} + e^{-i H_{\rm ho} t_0} \sigma^-_2  
e^{-ikx_2}\right]|\Psi(0)\rangle.
\ee
where $\sigma^x_1=\sigma^+_1+\sigma^-_1$.
After this interaction, the atoms continue their evolution with the free
harmonic oscillator Hamiltonian.

{\em (iii) State dependent kick of atom 2:} After a time $t_g-t_0$ a
short laser pulse is applied to atom $2$ with a k--vector pointing along
the $x$ direction as in the step (i) for a time $t_{\rm
las}=\pi/\Omega$. Assuming again $t_{\rm las}\ll t_g$, we can write the
state after this interaction as
\be
|\Psi(t_g)\rangle = 
(\sigma^+_2 e^{ikx_2} + \sigma^-_2 e^{-ikx_2})
e^{-i H(t_g-t_0)}
|\Psi(t_0)\rangle.
\ee
Using (\ref{period}) one can easily check that
\be
\label{Gate2}
|\Psi(t_g)\rangle = \left[ \sigma^+_2\sigma^-_2
+ \sigma^-_2\sigma^+_2 \sigma^x_1  \right] |\Psi(t_0)\rangle,
\ee
which coincides with the fundamental two--bit 
quantum (\ref{gate}).

Under ideal conditions the above steps allow to perform a two--bit
quantum gate. According to (\ref{Gate2}) the operators acting on the
motional state cancel out, and therefore the gate can be carried out
independent of the motional state, regardless of whether it is pure or
mixed. In a real situation, there will be restrictions to accomplish the
gate with high fidelity. The main sources of errors will be related to:
(a) the finite size of the wavepackets and their small separations, and
(b) the non--harmonic corrections to the free Hamiltonian (\ref{Hharm}).
The finite size effects can cause several problems during the step (ii):
(a.1) $\varphi^{R,L}$ may overlap at time $t_0$, which will prevent us
to address one of them alone with the laser beam; (a.2) even if the
wavepackets do not overlap, it will be hard (if not impossible) to focus
a laser beam to such small distances; (a.3) due to the spatial profile
of the laser beam different positions within the wavepacket will see
different laser intensities. On the other hand, if the wavepackets
separate from each other considerably, since this seems to be one way to
avoid some of the above mentioned problems, the anharmonic terms due to
$V_{\rm cor}$ may become important. In the following we will address all
these questions, find the conditions under which these problems can be
overcome, and present numerical simulations showing the performance of
the scheme in realistic setups.

Let us consider first the problems related to the finite size of the
atomic wavepackets and their small separations in step (ii). We denote
by $\Delta$ the size of the wavepackets corresponding to atom $1$ at
time $t_0$ and by $d$ their separation~(\ref{distance}) [Fig. 1(b)]. In
order to overcome the problem (a.1) it is required $D\gg \Delta$. The
laser profile is characterized by the position dependent Rabi frequency
$\Omega(x)$ which takes on a maximum value $\Omega_0$ and has a width
$W$. In present experiments, due to the impossibility of focusing laser
beams over small distances, this width is expected to be much larger
than the separation. Thus, 
\be
\label{Cond1}
W \gg D \gg \Delta. 
\ee
Consequently, the laser beam will affect both wavepackets $\varphi^{R,L}$,
which causes the problem (a.2). The solution to this problem is to
select the laser parameters so that $\varphi^{R,L}$ feel different Rabi
frequencies, $\Omega^{R,L}$, fulfilling 
\be
\label{Cond2}
\frac{\Omega^R}{2} t_1 = (2N+1/2)\pi,\quad \quad
\frac{\Omega^R}{2} t_1 = 2N\pi,
\ee 
with $N$ integer, the number of complete Rabi cycles. On the other hand,
we have to make sure that the whole wavepacket sees basically the same
Rabi frequency, so that no information of the internal state is
imprinted in the motional state, i.e. we have to overcome problem (a.3).
According to (\ref{Cond1}) this requires that 
\be
\label{Cond3}
\frac{d\Omega(x)}{dx}\Big|_{x=\overline{x}_1^{R,L}} \Delta \ll 1,
\ee
where $x=\overline{x}_1^{R,L}$ is the position of the center of the wave
packets of atom 1 when interacting with the laser. In order to
illustrate the above conditions, let us consider that the ions are
initially in a thermal state, characterized by $\overline{n}_c$, the
mean phonon number in the center--of--mass mode, and by $\overline{n}_r=
\overline{n}_c^2 / (2\overline{n}_c+1)$ the one in the relative motion.
For the laser profile we take a Gaussian $\Omega(x)=\Omega_0
\exp\left[-(x-l)^2/(2W^2)\right]$. We choose the equilibrium point of
atom $1$ to coincide with the steepest point of the laser profile, i.e.
$l=x_e/2 + W$. The condition (\ref{Cond1}) can now be expressed as $W
\gg 3\sqrt{3} x_0 \eta /2 \gg \sqrt{\overline{n}_c+\overline{n}_r/2+3/4}
x_0$ taking into account~(\ref{Dx1}). According to the first inequality,
we can expand the Gaussian profile $\Omega(x)$ around $x=x_e/2$ up to
first order. Imposing now the second condition (\ref{Cond2}) we obtain 
\be
\label{Cond2p}
\frac{\Omega(x_e/2)}{2} t_1 = (2N+1/4)\pi,
\ee 
and $W=(4N+1/2) D$. Having this in mind, the first condition (\ref{Cond1})
can be restated as
\be
\label{Cond1p}
4N\gg 1 ,\quad\quad \eta \gg \sqrt{4\overline{n}_c+2\overline{n}_r+3}/(3\sqrt{3}).
\ee
With this, condition (\ref{Cond3}) is automatically fulfilled. In
summary, the laser parameters have to be chosen following the conditions
(\ref{Cond2p}) and (\ref{Cond1p}). 

In order to illustrate to what degree the conditions derived above have
to be fulfilled, we have performed some numerical calculations. We have
considered as an example the potential $V(x)= K |x|^{5/3}$, which
ensures that $\nu_r=2\nu_c$. We have constructed the evolution according
to the harmonic approximation, and calculated the averaged fidelity
${\cal F}$ and purity ${\cal P}$ for different temperatures,
i.e. the mean center--of--mass phonon number $\overline{n}_c$, 
and values of $\eta$. As shown in Ref.\ \cite{Po97},
these quantities characterize the performance of the gate, and the
degree of decoherence, respectively. In Fig.\ 2 (a),(b) one can clearly see that
to obtain a good fidelity for high temperatures ($\overline{n}_c$) one has to
increase $\eta$ [see Eq.\ (\ref{Cond1p})]. We emphasize that if in a
given experiment $\eta$ is not large enough, one can simply apply a
sequence of $\pi$ pulses, from the left and right alternatively, to
increase the effective displacement of the wavepackets~\cite{We94}.
For example, the $S_{1/2} \rightarrow D_{5/2}$ transition of Ca$^+$ has
$\eta \approx 0.45$, for a trap frequency $\nu_c= 2 \pi \times 50$ kHz.
In order to obtain an effective value of $\eta=7$ one should apply a sequence 
of the order of 15 pulses.

We consider now the second source of errors, namely the effects of the
anharmonic term $V_{\rm cor}$ that we have neglected so far
(\ref{Htot}). In order to single out these effects, we will assume that
the the action of the laser beam in step (ii) is ideal. In that case,
the errors caused by anharmonicities are independent of the internal
dynamics, which simplifies the analysis. The fidelity of the gate will
be is then simply given by the overlap ${\cal F_{\rm cor}}=
|\langle\Psi(t_g)|\Psi(0)\rangle|^2$. As it should be, if we set $V_{\rm
cor}=0$ we will have ${\cal F_{\rm cor}}=1$ according to (\ref{period}).
Using time--dependent perturbation theory we find ${\cal F_{\rm cor}}=1-
(\Delta \tilde V)^2$, where
\be
\tilde V = \int_0^{t_g} d\tau e^{i H_{\rm ho} \tau}
V_{\rm cor} e^{-i H_{\rm ho} \tau}.
\ee
and $(\Delta \tilde V)^2 = \langle\Psi(0)|\tilde V^2|\Psi(0)\rangle
-  \langle\Psi(0)|\tilde V|\Psi(0)\rangle^2$. This fidelity can be 
evaluated analytically for an initial thermal state. In Fig.~2 (c) we have plotted 
${\cal F_{\rm cor}}$ as a function of $\eta$ and $\overline{n}_c$. Comparing
with Fig.~2 (a),(b), we see that the errors produced by anharmonicities can
be neglected with respect to the ones due to the finite size effects
of the wavepackets, at least for the values represented in these
plots. For larger values of $\eta$, however, the anharmonicity must
be taken into account. 

We have not considered the effects of decoherence during the gate
operations, such as thermalization of the phonon modes, spontaneous
emission {\em etc.}. The fundamental limits, however, will allow one to 
perform many gate operations during the decoherence time \cite{Ion,HughesKnight}.

In summary, we have shown how to implement two--bit quantum gates
between two ions in a linear trap at non--zero temperature. The scheme
can be easily generalized to the case of three ions. In contrast to the
ion trap proposal in \cite{Ci95} the present scheme is not easily scaled
up to more than three qubits, since one should tune the trap potentials
so that the frequencies of the different eigenmodes become
commensurable. We expect the present proposal to be of interest in
application of quantum logic with two and three qubits: examples are
fundamental experiments involving particle entanglement, demonstration
of error correction schemes \cite{St97}, and in quantum
communication \cite{Ci97,Va97}. 

J. F. P. acknowledges University of Innsbruck for their hospitality,
and a grant from The Junta de Comunidades de Castilla--La Mancha.
This research was supported in part by the Austrian Science Foundation
and the European TMR network ERB-FMRX-CT96-0087. 





\begin{references}


\bibitem{St97}
For a recent review see, for example,  A. Steane, report quant-ph/9708022 and
references therein.

\bibitem{Qc}
C. Monroe {\it et al}, Phys. Rev. Lett. {\bf 75}, 4714 (1995),
Q. A. Turchette {\it et al}, Phys. Rev. Lett. {\bf 75}, 4714 (1995);
X. Maitre, {\it et al}, Phys. Rev. Lett. {\bf 79}, 769 (1997).

\bibitem{Ci95}
J. I. Cirac and P. Zoller, Phys. Rev. Lett. {\bf 74}, 4091 (1995)

\bibitem{NMR}
N. A. Gershenfeld and I. L. Chuang, Science {\bf 275}, 350 (1996);
D. G. Cory, M. D. Price and T.  F.  Havel, report quant-ph/9709001 (1997).

\bibitem{Di89}
F. Diedrich {\it et al}, Phys. Rev. Lett. {\bf 62}, 403 (1989).


\bibitem{Po96}
For interferometry with ion traps see: J. F. Poyatos {\it et al}, Phys. Rev. A. {\bf 54}, 1532 (1996).

\bibitem{Ao}
For a review on atom optics see, for example, C. S. Adams, M. Sigel,
and J. Mlynek, Phys. Rep. {\bf 240}, 143 (1994).

\bibitem{Ion}
For theoretical and experimental reviews, see respectively
J. I. Cirac {\it et al}, Adv. At. Mol. Phys. {\bf 37}, 237 (1996),
D. J. Wineland {\it et al}, report quant-ph/9710025,
R. J. Hughes {\it et al}, report quant-ph/9708050. 

\bibitem{Note}
This can be accomplished using two co-propagating beams in a Raman transition,
or using a laser beam propagating in a direction perpendicular to the axis
$x$ for forbidden transitions.

\bibitem{Po97}
J. F. Poyatos, J. I. Cirac and P. Zoller, Phys. Rev. Lett. {\bf 78}, 390 (1997)

\bibitem{We94}
M. Weitz, B. C. Young and S. Chu, Phys. Rev. Lett. {\bf 73}, 2563 (1994)

%
\bibitem{HughesKnight}
R.J.~Hughes {\em et al.}, Phys. Rev. Lett. {\bf 77}, 3240 (1996);
M.B.~Plenio and P.L.~Knight, Phys. Rev. A {\bf 53}, 2986 (1996).

\bibitem{Ci97}
J. I. Cirac, {\it et al}, Phys. Rev. Lett. {\bf 78}, 3221 (1997)

\bibitem{Va97}
S. J. van Enk, J. I. Cirac and P. Zoller, Phys. Rev. Lett. {\bf 78},
4293 (1997);
S. J. van Enk, J. I. Cirac and P. Zoller (to be published).

\end{references}
\end{document}